\documentclass[preprint2]{aastex62}

\def\Mgii{Mg\,{\sc ii}}
\def\Civ{C\,{\sc iv}}
\def\Siiv{Si\,{\sc iv}}
\def\Nv{N\,{\sc v}}

\def\Feii{Fe\,{\sc ii}}

\def\Oiii{O\,{\sc iii}}

\shorttitle{A luminous BAL Quasar at $z=7.02$}
\shortauthors{Wang et al.}

\begin{document}

\title{The Discovery of A Luminous Broad Absorption Line Quasar at A Redshift of 7.02}

\correspondingauthor{Feige Wang}
\email{fgwang@physics.ucsb.edu}

\author[0000-0002-7633-431X]{Feige Wang}
\affil{Department of Physics, University of California, Santa Barbara, CA 93106-9530, USA}
\affil{Steward Observatory, University of Arizona, 933 North Cherry Avenue, Tucson, AZ 85721, USA}
\affil{Kavli Institute for Astronomy and Astrophysics, Peking University, Beijing 100871, China}

\author[0000-0001-5287-4242]{Jinyi Yang}
\affil{Steward Observatory, University of Arizona, 933 North Cherry Avenue, Tucson, AZ 85721, USA}

\author[0000-0003-3310-0131]{Xiaohui Fan}
\affil{Steward Observatory, University of Arizona, 933 North Cherry Avenue, Tucson, AZ 85721, USA}

\author[0000-0002-5367-8021]{Minghao Yue}
\affil{Steward Observatory, University of Arizona, 933 North Cherry Avenue, Tucson, AZ 85721, USA}

\author[0000-0002-7350-6913]{Xue-Bing Wu}
\affil{Kavli Institute for Astronomy and Astrophysics, Peking University, Beijing 100871, China}
\affil{Department of Astronomy, School of Physics, Peking University, Beijing 100871, China}

\author[0000-0002-4544-8242]{Jan-Torge Schindler}
\affil{Steward Observatory, University of Arizona, 933 North Cherry Avenue, Tucson, AZ 85721, USA}

\author[0000-0002-1620-0897]{Fuyan Bian}
\affil{European Southern Observatory, Alonso de C\'ordova 3107, Casilla 19001, Vitacura, Santiago 19, Chile}

\author[0000-0001-6239-3821]{Jiang-Tao Li}
\affil{Department of Astronomy, University of Michigan, 311 West Hall, 1085 South University Avenue, Ann Arbor, MI 48109-1107, USA}

\author[0000-0002-6822-2254]{Emanuele P. Farina}
\affil{Department of Physics, University of California, Santa Barbara, CA 93106-9530, USA}

\author[0000-0002-2931-7824]{Eduardo Ba\~nados}
\affil{The Observatories of the Carnegie Institution for Science, 813 Santa Barbara Street, Pasadena, CA 91101, USA}

\author[0000-0003-0821-3644]{Frederick B. Davies}
\affil{Department of Physics, University of California, Santa Barbara, CA 93106-9530, USA}

\author[0000-0002-2662-8803]{Roberto Decarli}
\affil{INAF--Osservatorio di Astrofisica e Scienza dello Spazio, via Gobetti 93/3, I-40129, Bologna, Italy}

\author{Richard Green}
\affil{Steward Observatory, University of Arizona, 933 North Cherry Avenue, Tucson, AZ 85721, USA}

\author[0000-0003-4176-6486]{Linhua Jiang}
\affil{Kavli Institute for Astronomy and Astrophysics, Peking University, Beijing 100871, China}

\author[0000-0002-7054-4332]{Joseph F. Hennawi}
\affil{Department of Physics, University of California, Santa Barbara, CA 93106-9530, USA}
\affil{Max Planck Institut f\"ur Astronomie, K\"onigstuhl 17, D-69117, Heidelberg, Germany}

\author[0000-0003-4955-5632]{Yun-Hsin Huang}
\affil{Steward Observatory, University of Arizona, 933 North Cherry Avenue, Tucson, AZ 85721, USA}

\author[0000-0002-5941-5214]{Chiara Mazzucchelli}
\affil{Max Planck Institut f\"ur Astronomie, K\"onigstuhl 17, D-69117, Heidelberg, Germany}

\author[0000-0002-3461-5228]{Ian D. McGreer}
\affil{Steward Observatory, University of Arizona, 933 North Cherry Avenue, Tucson, AZ 85721, USA}

\author[0000-0001-9024-8322]{Bram Venemans}
\affil{Max Planck Institut f\"ur Astronomie, K\"onigstuhl 17, D-69117, Heidelberg, Germany}

\author[0000-0003-4793-7880]{Fabian Walter}
\affil{Max Planck Institut f\"ur Astronomie, K\"onigstuhl 17, D-69117, Heidelberg, Germany}

\author{Yuri Beletsky}
\affil{Las Campanas Observatory, Carnegie Institution of Washington, Colina el Pino, Casilla 601, La Serena, Chile}

\begin{abstract}
Despite extensive efforts, to date only two quasars have been found at $z>7$, due to a combination of low spatial density and high contamination from more ubiquitous Galactic cool dwarfs in quasar selection. This limits our current knowledge of the super-massive black hole (SMBH) growth mechanism and reionization history. In this Letter, we report the discovery of a luminous quasar at $z=7.021$, DELS J003836.10--152723.6 (hereafter J0038--1527), selected using photometric data from DESI Legacy imaging Survey (DELS), Pan-STARRS1 (PS1) imaging Survey, as well as {\it Wide-field Infrared Survey Explore} ({\it WISE}) mid-infrared all-sky survey. With an absolute magnitude of $M_{1450}$=--27.1 and bolometric luminosity of $L_{\rm Bol}$=5.6$\times$10$^{13}$ $L_\odot$, J0038--1527 is the most luminous quasar known at $z>7$. Deep optical to near-infrared spectroscopic observations suggest that J0038--1527 hosts a 1.3 billion solar mass black hole accreting at the Eddington limit, with an Eddington ratio of 1.25$\pm$0.19. The \Civ\ broad emission line of J0038--1527 is blueshifted by more than 3000 km s$^{-1}$ relative to the quasar systemic redshift. More detailed investigations of the high-quality spectra reveal three extremely high-velocity \Civ\ broad absorption lines (BALs) with velocity from 0.08 to 0.14 times the speed of light and total ``balnicity" index of more than 5000 km s$^{-1}$, suggesting the presence of relativistic outflows. J0038--1527 is the first quasar found at the epoch of reionization (EoR) with such strong outflows, and therefore provides a unique laboratory to investigate active galactic nuclei feedback on the formation and growth of the most massive galaxies in the early universe.

\end{abstract}


\section{Introduction} \label{sec_intro}
As the most luminous non-transient objects, distant quasars are important tracers to study early structure formation and the history of cosmic reionization.
The detections of complete Gunn--Peterson (GP) absorption troughs in $z>6$ quasars mark the end of the reionization epoch at $z\gtrsim6$ \citep[e.g.,][]{fan06}. 
The Ly$\alpha$ damping wing absorption profile \citep{miralda98} probes neutral intergalactic medium (IGM) gas at the epoch of reionization (EoR).
Currently, only two luminous $z>7$ quasars have been reported in the literature \citep{mortlock11,banados18}, both of which exhibit signatures of strong damping wing absorption, suggesting that  the universe is significant neutral at $z>7$ \citep[e.g.,][]{bolton11,greig17,banados18,davies18}.

The existence of distant luminous quasars \citep{mortlock11, banados18} provides evidence of billion solar mass super-massive black holes (SMBHs) already formed in the EoR, which poses crucial constraints on their formation and growth mechanisms.
SMBH growth is not an isolated progress. It has been found to be tightly linked with the growth of their host galaxies \citep[i.e., the $M_{\rm BH}$-$\sigma_\ast$ relation;][]{gebhardt00}. Feedback by wind or outflow driven by black hole (BH) accretion has been invoked in simulations to explain the observed relation \citep[e.g.,][]{king03,dimatteo05}. Observationally, strong outflows are often studied in the rest-frame ultraviolet (UV) via blueshifted broad absorption lines \citep[BALs;][]{weymann91} or strong blue velocity shifts of broad \Civ\ emission lines in luminous quasars \citep[e.g.,][]{richards11}. These features appear most often at a moderate velocity of  $v<0.1$ times the speed of light ($c$), but relativistic BALs at $v\sim0.1-0.3c$ have been also found in a small number of quasars \citep[e.g.,][]{hamann13,rogerson16,hamann18}. The associated kinetic power of these relativistic outflows is estimated to be high enough to play a key role in the co-evolution of SMBHs and galaxies \citep[e.g.,][]{cicone15,feruglio17}, as it inevitably shocks against star formation and provide significant metal enrichment to the interstellar medium (ISM) and IGM \citep[e.g.,][]{chartas09,zubovas13}. 

However, to date only two $z>7$ quasars are known, which limits our understanding of the quasar population and their effects on galaxy formation and metal enrichment in the early Universe, highlighting the need to expand the sample of $z>7$ quasars.
In \cite{wang17}, we demonstrated that the combination of the Dark Energy Spectroscopic Instrument (DESI)\footnote{\url{http://desi.lbl.gov/}} Legacy Imaging Surveys \citep[DELS;][]{dey18} with near-infrared (NIR) surveys like the UKIRT Hemisphere Survey \citep[UHS;][]{dye18}, and the {\it Wide-field Infrared Survey Explore} mid-infrared survey \citep{wright10} allows us to search for the highest redshift quasars over a much large area than previous studies. Recently, we also updated our selection procedure by including the Pan-STARRS1 (PS1) Survey \citep{chambers16} to improve our selection efficiency \citep{wang18}.

Here, we report the discovery of a luminous $z>7$ BAL quasar, DELS J003836.10--152723.6 (hereafter J0038--1527), from our ongoing distant quasar survey. 
In \S\ \ref{sec_obs}, we present our spectroscopic observations and infrared photometric observations. In \S\ \ref{sec_bh}, we describe the luminosity and BH mass measurements. In \S\ \ref{sec_outflow}, we characterize the strong relativistic outflows detected in this quasar. Finally, in \S\ \ref{sec_discussion} we briefly discuss the implications and summarize future investigations on high-redshift quasars. 
All results below refer to a $\Lambda$CDM cosmology model with a Hubble parameter of $H_0=70$ km s$^{-1}$ Mpc$^{-1}$ and density parameters of $\Omega_m=0.3$, and $\Omega_\Lambda=0.7$.

\section{Observations and Data Analysis} \label{sec_obs}
J0038--1527 was selected as a $z>6.5$ quasar candidate based on DELS and PS1 photometric data. It is detected in both DELS $z$-band ($z_{\rm AB}$ = 21.65$\pm$0.08), and PS1 $y$-band ($y_{\rm AB}$ = 20.61$\pm$0.10), but is undetected in PS1 $g$, $r$, $i$, and $z$ bands. The strong dropout nature makes it is a promising high-redshift quasar candidate. 
J0038--1527 is also detected in ALLWISE ($W1_{\rm VEGA}$ = 16.80$\pm$0.10, $W2_{\rm VEGA}$ = 16.08$\pm$ 0.21) and therefore it is unlikely to be a nearby Galactic cool dwarf with high proper motion.
J0038-1527 is one of the brightest candidates in our sample, and thus has high priority when taking spectroscopic follow-up observations. We refer to \cite{wang17} and \cite{wang18} for detailed descriptions of the target selection. 

The initial spectroscopic observation was obtained on 2018 January 16  with Multiple Mirror Telescope (MMT)/Red Channel Spectrograph using a 270 mm$^{-1}$ grating and 1\farcs25 slit, which provide a spectral resolution of $R\sim$500. The spectrum shows a clear break at $\sim$9700\AA\ , which suggests it is a quasar at $z\sim7$.  Subsequent high signal-to-noise ratio (S/N) spectra taken with the MMT and Magellan/LDSS3-C confirm it as a quasar at $z>7$. 

We were allocated Very Large Telescope (VLT)/X-SHOOTER DD time (program ID: 2100.A-5033(A)) and the observations were obtained over four nights between 2018 January and July. The total on-source exposure was 12,000s. 
We used 0\farcs9 slits in both optical (VIS) and NIR arms, which deliver resolutions of $R\sim8900$ and $R\sim5600$ in VIS and NIR arms, respectively. The data were reduced using standard European Southern Observatory (ESO) X-SHOOTER pipeline.

\begin{figure*}
\centering
\includegraphics[width=1.0\textwidth]{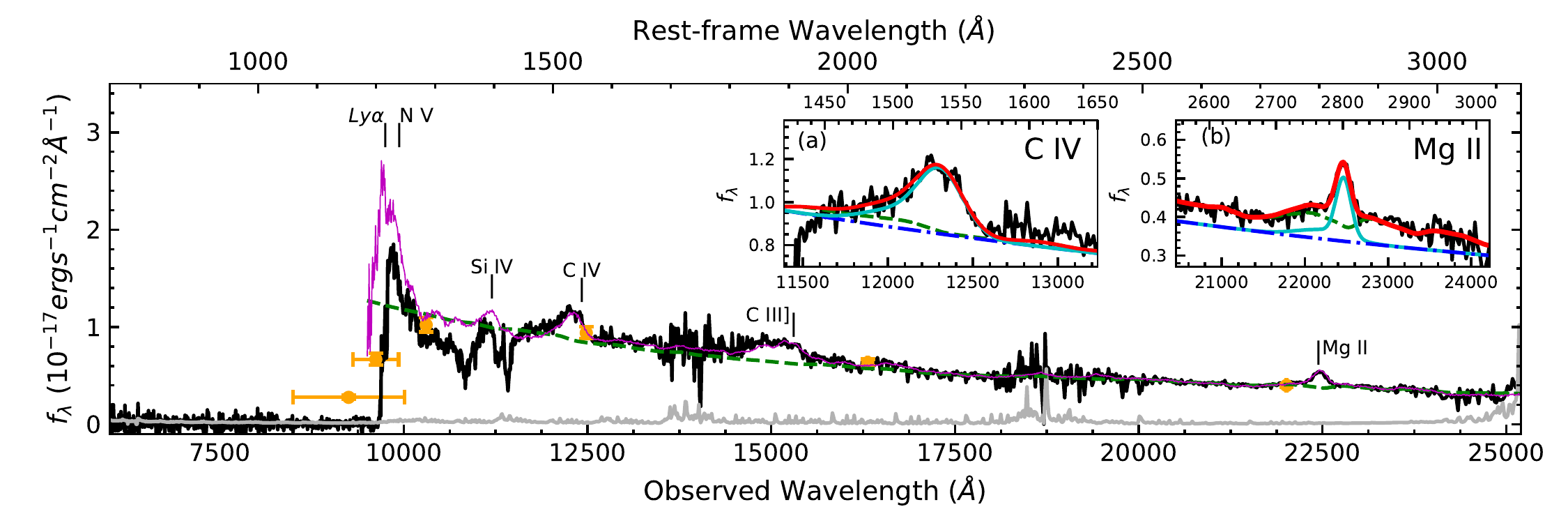}
\caption{Final calibrated spectrum of J0038--1527. The black and gray lines represent the Galactic extinction corrected spectrum and error vector. The thin magenta line denotes the quasar composite spectrum constructed with $\sim200$ SDSS quasars with large \Civ\ blueshifts. The green dashed line denotes the pseudo-continuum which includes power-law,  \Feii\ emission, and Balmer continuum . The orange circles are flux density converted from galactic extinction corrected magnitudes listed in Table \ref{phot}. The x-axis error-bars of the two leftmost orange circles denote the FWHMs of $z_{\rm DELS}$ and $y_{\rm ps1}$ filter curves.
Two inner plots show spectral fitting of \Civ\ (a) and \Mgii\ (b) regions, respectively. The blue dot-dashed line denotes the best-fit power-law continuum, the green dashed line denotes the best-fit pseudo-continuum. The cyan line denotes fitted \Civ\ and \Mgii\ emission lines plus power-law continuum and the red line denotes the total fitted flux. 
\label{spec}}
\end{figure*}

Additional NIR spectroscopy was obtained with Gemini/GNIRS, with a total exposure time of 4.2 hr (program ID: GN-2018A-FT-114). We used  the cross-dispersed mode with a 0\farcs675 slit, which provides a resolution of $R\sim$750 from $\sim$0.9$\mu$m to $\sim$2.5$\mu$m. The GNIRS data was reduced with the XIDL\footnote{\url{https://github.com/profxj/xidl}} suite of astronomical routines in the Interactive Data Language (IDL).
Finally, we produced the combined spectrum using the X-SHOOTER and GNIRS observations and 
scaled it to match the photometric data for absolute flux calibration. Then we correct the Galactic extinction using the \cite{cardelli89} reddening law and $E(B-V)$ from \cite{schlegel98}. The final calibrated optical to NIR spectrum is shown in Figure \ref{spec}.

In order to constrain the rest-frame UV spectral energy distribution (SED) of J0038--1527, we obtained Y, J, H, and K-band photometry using UKIRT/WFCam (Project ID: U/17B/D04) on 2018 January 20. The on-source exposures were 12 mins in the Y, J, and K-bands and 6 minutes in the H-band. The processed data was kindly provided by M. Irwin using the standard Visible and Infrared Survey Telescope for Astronomy (VISTA)/WFCAM data-flow system \citep{irwin04}. The photometric properties and derived parameters of J0038--1527 are listed in Table \ref{phot}. 

\begin{deluxetable}{lc}
\tabletypesize{\scriptsize}
\tablecaption{Photometric Properties and Derived Parameters of J0038--1527. \label{phot}}
\tablewidth{0pt}

\startdata
\\
R.A. (J2000) & 00:38:36.10\\
Decl. (J2000) & --15:27:23.6 \\
Redshift   & 7.021$\pm$0.005 \\
$m_{1450}$ & 19.84$\pm$0.08 \\
$M_{1450}$ & -27.10$\pm$0.08 \\
$z_{\rm DELS, AB}$ & 21.65$\pm$ 0.08\\
$y_{\rm ps1, AB}$ & 20.61$\pm$ 0.10 \\
$Y_{\rm VEGA}$   & 19.39$\pm$ 0.08\\
$J_{\rm VEGA}$   & 18.75$\pm$ 0.07\\
$H_{\rm VEGA}$   & 18.15$\pm$ 0.06\\
$K_{\rm VEGA}$   & 17.48$\pm$ 0.05\\
$W1_{\rm VEGA}$   & 16.80$\pm$ 0.10\\
$W2_{\rm VEGA}$   & 16.08$\pm$ 0.21\\
$g_{\rm ps1},r_{\rm ps1},i_{\rm ps1},z_{\rm ps1}$\tablenotemark{a} & $>24.1,24.0,23.7,22.9$ \\
\hline
$z$$\rm _{MgII}$ & 7.025$\pm$0.005\\
$z$$\rm _{CIV}$ & 6.939$\pm$0.008\\
$\alpha_\lambda$ & --1.54$\pm$0.05\\
$\rm \Delta_{v_{CIV}-v_{MgII}}$  (km s$^{-1}$) & 3400$\pm$411\\
FWHM$\rm _{MgII}$ (km s$^{-1}$) & 2994$\pm$140\\
EW$\rm _{MgII}$ (\AA) & 16.5$\pm$1.0\\
FWHM$\rm _{CIV}$ (km s$^{-1}$) & 8728$\pm$452\\
EW$\rm _{CIV}$ (\AA) & 18.1$\pm$1.4\\
$\lambda L_{\rm 3000\text{\normalfont\AA}}$ (erg s$^{-1}$) & 4.19$\times$$10^{46}$\\
$L_{\rm Bol}$ (erg s$^{-1}$) & 2.16$\times$$10^{47}$\\
$M_{\rm BH}$ ($\rm M_\odot$) & (1.33$\pm$0.25)$\times$10$^9$\\
$L_{\rm Bol}/L_{\rm Edd}$ & 1.25$\pm$0.19\\
\enddata
 \tablenotetext{a}{Magnitude limits at 3-$\sigma$ level.}
\end{deluxetable}

\section{Luminosity and BH Mass} \label{sec_bh}
\begin{figure}
\centering
\includegraphics[width=0.5\textwidth]{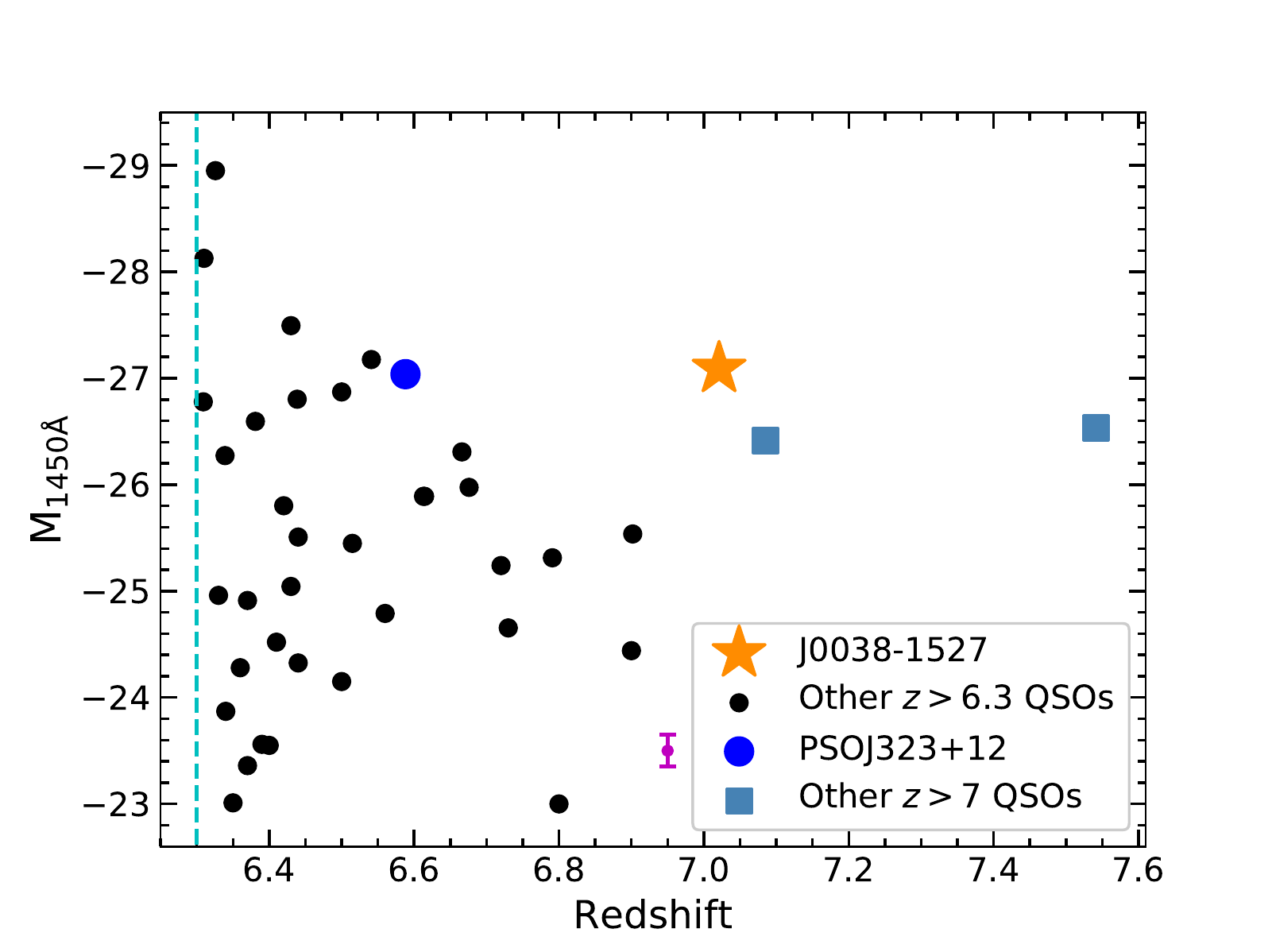}
\caption{Absolute magnitude of all publicly known $z>6.3$ quasars. J0038--1527 with $M_{1450,AB}$=--27.10, is the most luminous $z>6.6$ quasar known to date. The next-highest redshift quasar (PSO323+12 at $z=6.588$) with comparable luminosity is highlighted by a large blue circle. Two previously known $z>7$ quasars are denoted by squares. The magenta error bar shows the typical errors on $M_{1450,AB}$ of publicly known quasars. The cyan dashed line mark the position of $z=6.3$. \label{fig_mz}}
\end{figure}

We fit the final calibrated spectrum following the approach detailed in \cite{wang15}. 
Briefly, we shift the spectrum to rest frame using an initial redshift, and fit a pseudo-continuum model that includes a power-law continuum, \Feii\ emission \citep{vestergaard01,tsuzuki06}, and Balmer continuum \citep[e.g.,][]{derosa14} to the line-free regions using a $1/\sigma^2$ weighting $\chi^2$ fitting technique. We then fit the \Mgii\ emission and derive the quasar systemic redshift by correcting \Mgii\ redshift to [\Oiii] redshift based on the velocity offset between these two lines of the SDSS quasar composite \citep{van01}. We iterated this procedure until the difference of input redshift and output redshift is smaller than the uncertainty. Then the spectrum is de-redshifted using the final systemic redshift, which is $z=7.021\pm0.005$.
The final fitting yields a power-law continuum of $f_\lambda \propto \lambda^{-1.54\pm0.05}$. 
We measure the rest-frame 3000 \AA\ power-law luminosity to be $\lambda L_{\rm3000\text{\normalfont\AA}}$=4.19$\times$$10^{46}$ erg s$^{-1}$, and  
the rest-frame 1450 \AA\ magnitude to be $m_{1450,AB}$=$19.84\pm0.08$ and $M_{1450,AB}$=--27.10$\pm$0.08. J0038--1527 is about a half magnitude brighter than the other two known $z>7$ quasars and is the only known quasar at $z>6.6$ with $M_{1450,AB}<-27$ (Figure \ref{fig_mz}). By assuming an empirical conversion factor from the luminosity at 3000 \AA\ \citep[e.g.,][]{shen11}, we estimate the bolometric luminosity of J0038--1527 as $L_{\rm bol}$=5.15$\times$ $\lambda L_{\rm3000 \text{\normalfont\AA}}$ = 2.16$\times$10$^{47}$ erg s$^{-1}$ = 5.6$\times$10$^{13}$ $L_\odot$. 

After subtracting the best-fit pseudo-continuum from the spectrum, we fit \Mgii\ and \Civ\ broad emission lines with two Gaussian profiles for each line. We measure a full-width at half-maximum (FWHM) of $2994\pm140$ km s$^{-1}$ and $8728\pm452$ km s$^{-1}$ for \Mgii\ and \Civ, respectively. The continuum and line fitting are shown in Figure \ref{spec}. After applying a virial BH mass estimator \citep{vestergaard09} based on the \Mgii\ line, we estimate the SMBH mass of J0038--1527 to be (1.33$\pm$0.25)$\times$10$^{9}$ $\rm M_\odot$. The accretion rate of this  quasar is consistent with Eddington accretion, with an Eddington ratio of $L_{\rm Bol}/L_{\rm Edd}$=1.25$\pm$0.19. Note that the Eddington ratio quoted here depends on the virial BH mass estimator that we adopt. Using the relation from \cite{mclure04}, we get an SMBH of (1.21$\pm$0.36)$\times$10$^{9}$ $\rm M_\odot$ and $L_{\rm Bol}/L_{\rm Edd}$=1.37. The quoted uncertainty does not include the systematic uncertainties in the scaling relation, which could be up to $\sim0.4$ dex \citep{shen11}. 

The BH growth in the early universe is limited by the available accretion time. 
The existence of J0038--1527 and the other two $z>7$ quasars requires either massive seed BHs or episodes of super-Eddington accretion under a typical radiation efficiency of $\epsilon \sim$ 0.1 to form billion solar mass BHs in such young Universe. Otherwise, the radiation efficiency must be much lower than that allowed for thin disk accretion.

\begin{figure}[tbh]
\centering
\includegraphics[width=0.5\textwidth]{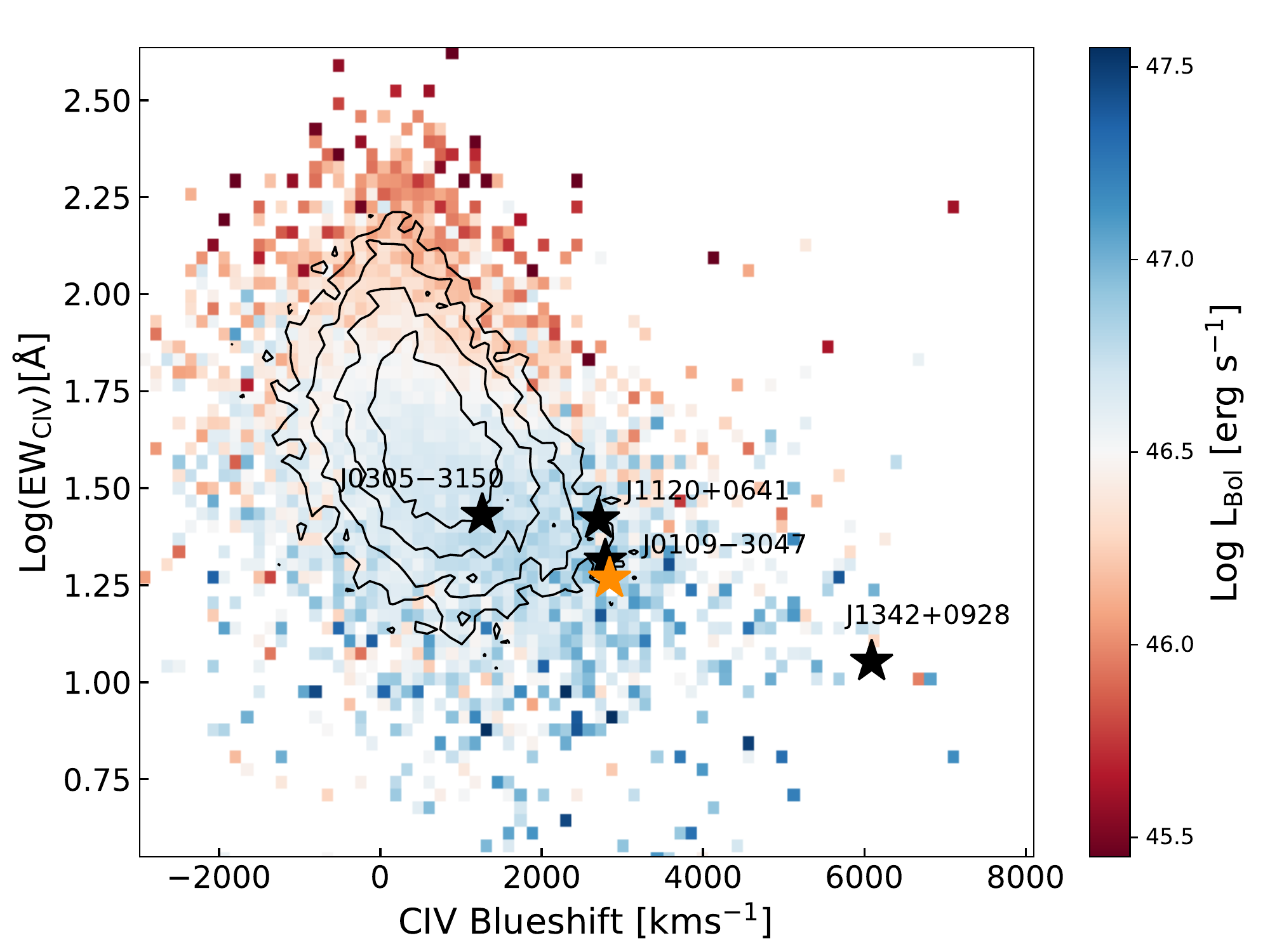}
\caption{\Civ\ emission line blueshift vs. \Civ\ EW and quasar bolometric luminosity. The contours and 2D histogram are for SDSS low-redshift quasars \citep{shen11} and the black asterisks denote public known $z\gtrsim6.5$ quasars with \Civ\ observations \citep[e.g.,][]{mortlock11,derosa14,mazzucchelli17,banados18}. The orange asterisk represents J0038--1527. Although all $z\gtrsim6.5$ quasars are in the large \Civ\ blueshift tail, they are still consistent with the distribution of low-redshift quasars of similarly high luminosities.
\label{fig_voff}}
\end{figure}

\section{Quasar Outflows} \label{sec_outflow}
As described in \S \ref{sec_intro}, quasar outflows are thought to play an important role in regulating the co-evolution of  central SMBHs and host galaxies.
Both broad emission lines like \Civ\ and highly ionized absorption lines can be efficient diagnostics of quasar outflows. 
Especially, high-ionization broad emission lines like \Civ\ of quasars are usually blueshifted from the systemic velocity by several hundred km s$^{-1}$ and up to $\sim$7000 km s$^{-1}$, depending on the equivalent width (EW) of \Civ\ and the quasar luminosity \citep[e.g.,][also Figure \ref{fig_voff}]{richards11}. The anti-correlation between continuum luminosity and emission line EW is well known as the Baldwin Effect \citep{baldwin77}. 
More recently, results from the SDSS quasar reverberation mapping project \citep{sun18} show that 
extreme blueshift quasars have a low level of variability, suggesting that high-blueshift sources tend to also have high Eddington ratios. 
From our analysis of J0038--1527, we find that the peak of the \Civ\ emission has a strong blueshift with velocity of 3400$\pm$411 km s$^{-1}$ compared to \Mgii. We measured the rest-frame EWs to be EW$_{\rm CIV}$=$18.1\pm1.4$ \AA\ and EW$_{\rm MgII}$=$16.5\pm1.0$ \AA. This places J0038--1527 at the high-blueshift velocity end of \Civ\ emission line. 
In order to further confirm this, we used a more robust way to measure the \Civ\ blueshift explored by \cite{coatman16,coatman17}, which measures the line centroid as the wavelength that bisects the cumulative line flux. This method yields a \Civ\ blueshift of 3800 km s$^{-1}$, confirms the strong blueshifted \Civ\ emission line in J0038--1527.

\begin{figure}
\centering
\includegraphics[width=0.5\textwidth]{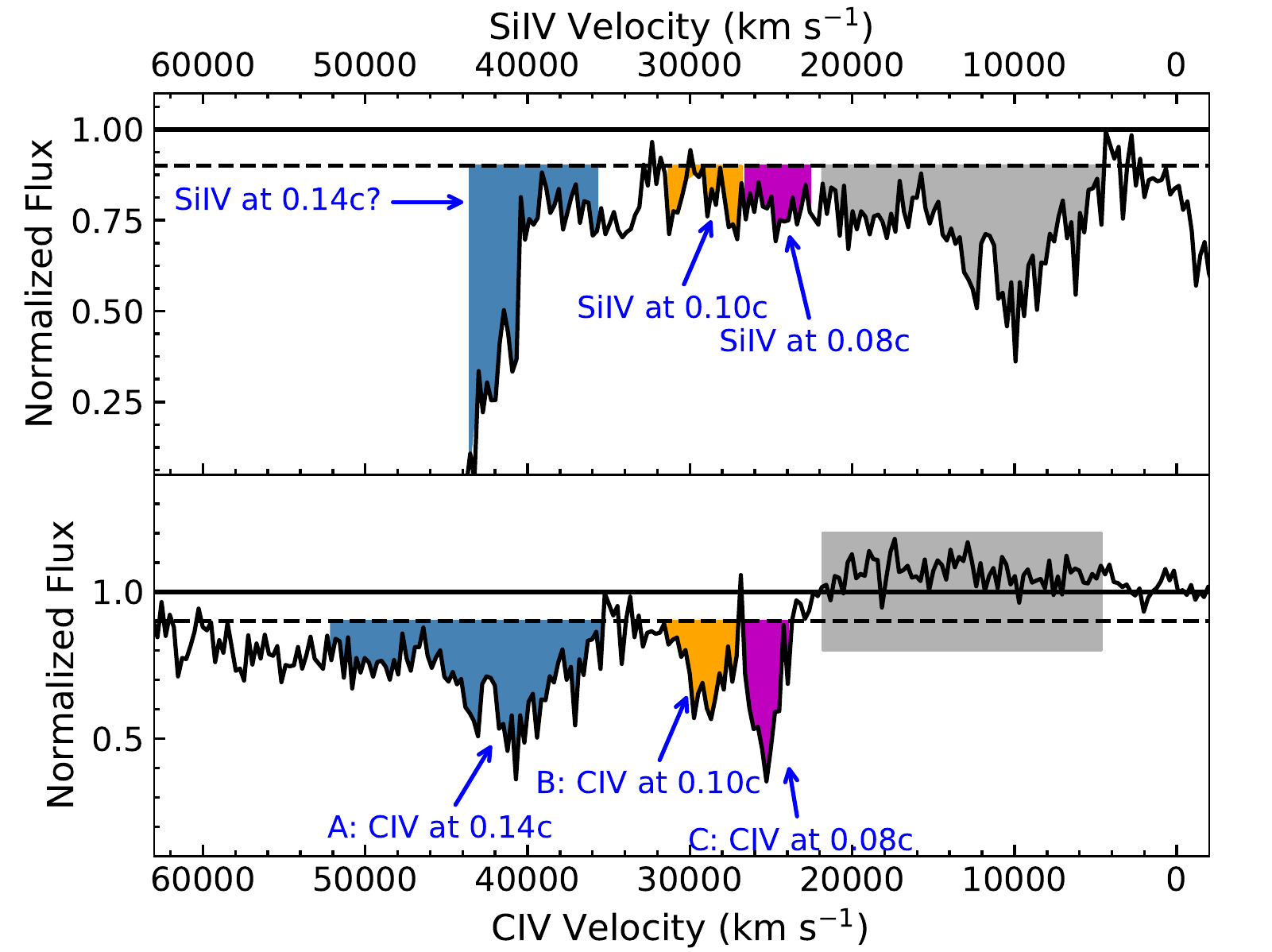}
\caption{Normalized spectrum of J0038--1527. The x-axis of the top panel is the outflow velocity of \Siiv\ and the x-axis of bottom panel shows the outflow velocity relative to \Civ. The blue, orange and magenta shaded regions denote three absorption systems at 0.14$c$ (A), 0.10$c$ (B), and 0.08$c$ (C), respectively. The velocity range of each trough was determined by the \Civ\ trough. The absence of absorption in the gray shaded region in the bottom panel and strong absorption on top of Ly$\alpha$ and \Nv\ (blue shaded region in the top panel) suggests that the trough A is indeed an extremely high velocity \Civ\ BAL.
\label{fig_bal}}
\end{figure}

More interestingly, after examining the spectrum of J0038--1527, we find several strong BAL features blueward of the \Civ\ and \Siiv\ emission lines. No absorption feature is found blueward of \Mgii\, indicating that it is a high-ionization broad absorption line quasar (HiBAL).
In order to estimate the intrinsic spectrum of J0038--1527, we construct a quasar composite spectrum (Figure \ref{spec}) using $\sim$200 SDSS quasars with extreme \Civ\ emission line blueshifts ($>3300$ km s$^{-1}$). We then divide the spectrum of J0038--1527 by the matched composite to derive a normalized spectrum.
From the normalized spectrum, we identified three \Civ\ absorption troughs at extremely high velocities of $(0.14_{-0.02}^{+0.03}) c$ (trough A), $(0.10_{-0.01}^{+0.005})c$ (trough B), and $(0.08_{-0.005}^{+0.008})c$ (trough C), which are highlighted with blue, orange, and magenta shaded regions in Figure \ref{fig_bal}. These three troughs also have accompanied \Siiv\ troughs (top panel in Figure \ref{fig_bal}). In order to quantify the strength of these troughs, we measure the ``balnicity" index (BI, \cite{weymann91}) of \Civ\ BALs by
\begin{equation}
{\rm BI} = \int_{v_{\rm min}}^{v_{\rm max}}\left ( 1 - \frac{f(v)}{0.9} \right )C dv.
\end{equation}
where $f(v)$ is normalized spectrum, $C$ is set to 1 only when $f(v)$ is continuously smaller than 0.9 for more than 2000 km s$^{-1}$, otherwise it is set to 0.0. 
In order to avoid counting absorptions from  \Siiv, we set $v_{\rm max}=52,340$ km s$^{-1}$. The $v_{\rm min}$ is set to 0.
The BIs are measured to be 3400 km s$^{-1}$, 890 km s$^{-1}$, and 1010 km s$^{-1}$ for trough A, B, and C, respectively. 
The uncertainties of the BIs are dominated by the limitations of the template matching accuracy.
The total \Civ\ BI of J0038--1527 is 5300 km s$^{-1}$, which is on the high tail ($\sim$10\%) of the BI distribution of BAL quasars at lower redshifts \citep[e.g.,][]{gibson09}.

We note that the composite spectrum matches the spectrum of J0038--1527 very well from \Civ\ to \Mgii\ but slightly underestimates the continuum blueward of \Civ\ broad emission line. Such difference could be caused by the fact that high-redshift quasars tend to have flatter extinction curves than that of low-redshift quasars \citep[e.g.,][]{maiolino04, gallerani10}. If so, we might slightly underestimate the BIs measured above. 
Because trough A has a very high velocity, the associated \Siiv\ absorption trough is blueshifted to the Ly$\alpha$ region, where the spectrum is seriously absorbed by the intervening IGM. On the other hand, the best-fit composite template clearly overfits \Nv\, and maybe also Ly$\alpha$, suggesting strong absorptions over this region, which could be due to \Siiv\ BAL at $\sim0.14c$.
The other possibility for trough A is that it could be a \Siiv\ BAL trough at a lower velocity, as this trough is at the blueward of \Siiv\ emission line. However, we do not see any accompanied \Civ\ troughs at the same velocity shifts (gray shaded region in Figure \ref{fig_bal}). Without associated \Civ\ troughs, the presence of strong absorption on the top of Ly$\alpha$ and \Nv\ emission lines, and the existence of the other two high-velocity troughs (B and C), the spectrum of J0038--1527 strongly suggests that trough A is indeed an extremely high-velocity \Civ\ BAL. 
\cite{allen11} found that the BAL quasar fraction increases by a factor of 3.5 from  $z\sim2$ to $z\sim4$, which indicates that an orientation effect alone is not sufficient to explain the presence of BAL troughs. This, together with the strong BAL absorptions in J0038--1527 and large \Civ\ blueshifts found in those $z\gtrsim7$ quasars, suggests that strong outflows are common in the earliest quasars.

\section{Discussion and Summary} \label{sec_discussion}

As described in \S \ref{sec_outflow}, although most $z>6.5$ quasars show strong \Civ\ emission line blueshifts, they still follow the locus of low-redshift quasars in Figure \ref{fig_voff}. Moreover, a low-redshift quasar composite constructed using a simple cut of \Civ\ blueshifts matches the spectrum of J0038--1527 very well.  This suggests that the property of \Civ\ emission line is very important for damping wing analysis, as the main uncertainty on such analysis comes from how well one can predict the intrinsic spectra of high-redshift quasars. 

The relativistic outflows in this quasar are found with extremely high velocities of 0.08$c$--0.14$c$. Such phenomenon is very rare and only observed in a small number of low-redshift quasars \citep[e.g.,][]{rogerson16, hamann18}. Very recently, \cite{hamann18} found that highly ionized X-ray ultra-fast outflow \citep[UFO;][]{chartas02,reeves03} in a low-redshift quasar is accompanied with a high-velocity \Civ\ BAL, which suggests that the relativistic \Civ\ BALs might form in the dense clumps embedded in the X-ray UFO. The associated kinetic power of these relativistic outflows is suggested to be well above what's needed to affect quasar host galaxies \citep[e.g.,][]{chartas09,zubovas13}. J0038--1528 is an excellent target to test whether or not AGN feedback could affect the building up progress of massive galaxy with future deep X-ray observations \citep[e.g.,][]{chartas09} and (sub-)millimeter observations \citep[e.g.,][]{feruglio17}. The most distant quasars may have more ubiquitous and stronger outflows of dense gas than their low-redshift counterparts \citep[see also][]{maiolino04}. Future larger sample of quasars at $z\gtrsim7$ are needed to measure the BAL quasar fraction and characterize outflow properties in a statistical manner.  

With $\rm M_{1450}$=--27.1, J0038--1527 is the most luminous $z>7$ quasar known to date. The BH mass of J0038--1527 is measured to be $M_{\rm BH}$=(1.33$\pm$0.25) $\times$ 10$^9$ $\rm M_\odot$ based on \Mgii\ emission line. 
We estimated the bolometric luminosity of J0038--1527 to be  $L_{\rm Bol}$=2.16$\times$$10^{47}$ erg s$^{-1}$, which yields an Eddington ratio to be $L_{\rm Bol}/L_{\rm Edd}$ = 1.25$\pm$0.19, suggesting a rapid BH growth phase. 
The temperature of the accretion disk increases with the accretion rate, and thus produces more UV photons. At the same time, the inner accretion disk is puffed up and could act as a ``shielding'' gas \citep[e.g.,][]{abramowicz88} to block the X-ray emissions. Moreover, the inverse Compton scattering of UV photons cools the hard X-ray corona emission more efficiently resulting in a soft SED \citep[e.g.,][]{jiang14a}. Quasars with soft SEDs can launch strong winds from the accretion disk \citep[e.g.,][]{jiang14b}, which could produce the observed strong blueshifts of \Civ\ broad emission lines and BAL features \citep[see][and references therein]{richards11}. Thus, the observed outflows indicated by both \Civ\ emission line blueshifts and BALs are probably driven by high Eddington ratios observed in luminous quasars as suggested by \cite{sun18}, consistent with the properties observed in J0038--1527.

\acknowledgments
J. Yang, X. Fan, M. Yue, J.-T. Schindler and I. D. McGreer acknowledge support from the US NSF Grant AST-1515115 and NASA ADAP Grant NNX17AF28G. X.-B.W. and L.J. acknowledge support from the National Key R\&D Program of China (2016YFA0400703) and the National Science Foundation of China (11533001 \& 11721303). B. Venemans and F. Walter acknowledge funding through the ERC grants ``Cosmic Dawn" and ``Cosmic Gas". 
This research based on observations obtained at the Gemini Observatory (GN-2018A-FT-114) and based on observations collected at the European Organization for Astronomical Research in the Southern Hemisphere under ESO program 2100.A-5033 (A). We especially thank the Directors of VLT, and UKIRT for granting us Director Discretionary time for follow up observations of this object. We acknowledge the use of data obtained at the Gemini Observatory, Magellan Telescope, and MMT Observatory. We acknowledge the use of public DELS, PS1, and {\it WISE} data. 

\vspace{5mm}
\facilities{Gemini (GNIRS), Magellan (LDSS3-C), MMT (Red Channel Spectrograph), UKIRT (WFCAM), VLT (X-SHOOTER)}

\end{document}